# DETECTING THE HISTORICAL ROOTS OF TRIBOLOGY RESEARCH: A BIBLIOMETRIC ANALYSIS

Bakthavachalam Elango[1*], Lutz Bornmann[2] and Govindaraju Kannan[3]

[1]Library,
IFET College of Engineering,
Villupuram – 605 108,
India.
Email: elangokb@yahoo.com

[2]Division for Science and Innovation Studies,
Administrative Headquarters of the Max Planck Society,
Hofgartenstr. 8,
80539 Munich, Germany.
Email: bornmann@gv.mpg.de

[3]Department of Mechanical Engineering,
IFET College of Engineering,
Villupuram – 605 108,
India.
Email: khannanvignesh@yahoo.co.in

*Corresponding Author

**ABSTRACT**

In this study, the historical roots of tribology (a sub-field of mechanical engineering and materials science) are investigated using a newly developed scientometric method called "Reference Publication Year Spectroscopy (RPYS)". The study is based on cited references (n=577,472) in tribology research publications (n=24,086). The Science Citation Index – Expanded (SCI-E) is used as data source. The results show that RPYS has the potential to identify the important publications: Most of the publications which have been identified in this study as highly cited (referenced) publications are landmark publications in the field of tribology.

**Keywords**: Bibliometrics, Citation Peak, Reference Publication Year Spectroscopy (RPYS), Historical Roots, Tribology

**INTRODUCTION**

New research usually evolves on the basis of previous investigations and discussions among the experts in a specific scientific community. Although there are many differences between the theories about scientific development (see e.g. Popper, 1961 and Kuhn, 1962), the relationship of current research to past literature always plays a significant role: knowledge cannot be acquired without the references to the past (Bornmann, de Moya Anegón & Leydesdorff 2010). Although the past literature plays a significant role in every research field, their importance is seldom studied using scientometric techniques and data. Thus, Marx et al. (2014) introduced the method "Reference Publication Year Spectroscopy (RPYS)" to reveal the important historical publications in a research field. The RPYS is able to identify the historical roots of research fields and can quantify their citation impact on current research. The method is based on analyzing the frequencies with which references are cited in the publications of a specific research field in terms of the publication years of the cited references. According to Marx and Bornmann (2014), RPYS can not only be applied to the identification of historical roots, but also to unveil scientific legends in a scientific field.

This study is intended to identify the historical roots of the tribology research from the perspective of the cited references. The term *tribology* was coined by Jost (1966) deriving from the Greek word *tribos* (or *triben*) means rubbing. Tribology is the science and technology of two interacting surfaces in relative motion and of related subjects and practices. Tribology is a multidisciplinary field which incorporates a number of disciplines, including mechanical engineering, material science, mechanics, surface chemistry, and surface physics. According to a report of the South African Institute of Tribology, tribology is a property of matter or the second most important field of study of material property after that of gravity.

**METHODOLOGY**

The results of the RPYS for the tribology field is based on the Science Citation Index - Expanded (SCI-E, Thomson Reuters). The study is mainly concerned with the analysis of the reference publication years (RPYs) and especially with the analysis of early publications cited particularly frequently as the historical roots of tribology research. In order to analyze the RPYs, the following steps have been employed with the program *rpys.exe* (see http://www.leydesdorff.net/software/rpys/ and Bornmann et al. in press ).

- The $1^{st}$ step is to select the publications on tribology in the SCI-E database and to extract all bibliographical records.
- The $2^{nd}$ step is to extract all references from the records using *rpys.exe* and to identify the most important historical RPYs for the tribology research field.
- The $3^{rd}$ step is to identify the most important publications in specific RPYs using the program *yearcr.exe*. The program *RefMatchCluster.jar* has been employed to aggregate cited references across misspellings and variants.
- The $4^{th}$ step is to establish the frequency distribution of cited references over the RPYs and to determine the publications cited most frequently in early RPYs.

The publications on tribology were selected in the SCI-E database by searching in the title, abstract, author keywords and keywords plus fields with the following keywords (date of search: May 2015): *\*tribolog\* OR "tribosyst\*" OR "tribo-syst\*" OR "tribo-chem\*" OR "tribochem\*" OR "tribotechn\*" OR "tribo-physi\*" OR "tribophysi\*"* (Elango et al. 2015; Elango et al. in press). An overview of the data set used in this study is provided in Table 1.

| Table 1 – General overview of the data set used | |
|---|---|
| **Item** | **#** |
| Number of publications | 24086 |
| Period of publication | 1953-2014 |
| Number of cited references | 577472 |

Based on the SCI-E input data (publications on tribology), *rpys.exe* generates two output files: *rpys.dbf* contains the number of cited references per RPY. *median.dbf* contains the deviation of the number of cited references in each RPY from the median for the number of cited references in the two previous, the current, and the two following RPYs [t - 2; t - 1; t; t + 1; t + 2]. Both files are used in Excel for drawing a spectrum (see Figure 1) and heat map (see Figure 2).

Bornmann et al. (in press) recommend to calculate quantile values in order to compare the importance of different RPYs. For the calculation, the formula given by Hazen (1914) is employed:

$$\text{Quantile} = ((i-0.5)/n * 100),$$

Where $i$ is the rank of a specific RPY (years are ranked in decreasing order by their number of cited references) and $n$ the total number of RPYs. The quantile values are available in *median.dbf* generated by *rpys.exe*. The higher the quantile value for a specific RPY, the most frequently referenced literature (cited) from that RPY (compared to other RPYs).

**RESULTS**

The distribution of the number of references cited in the tribology literature is presented in Figure 1. The most frequently cited RPY is 2000, showing the strong contemporary relevance of this research field. Figure 2 shows the heat map based on quantiles for the RPYs. The figure reveals that the most frequently cited RPYs spread between 1999 and 2006. However, some RPYs in the early years seem to be important too (e.g. 1805 or 1882). In order to receive an overview of the history in tribology, we limited the RPYs to the period between 1801 and 1965. The term "tribology" was introduced by Jost (1966), so that the year 1966 which might be seen as the starting point of the modern tribology research.

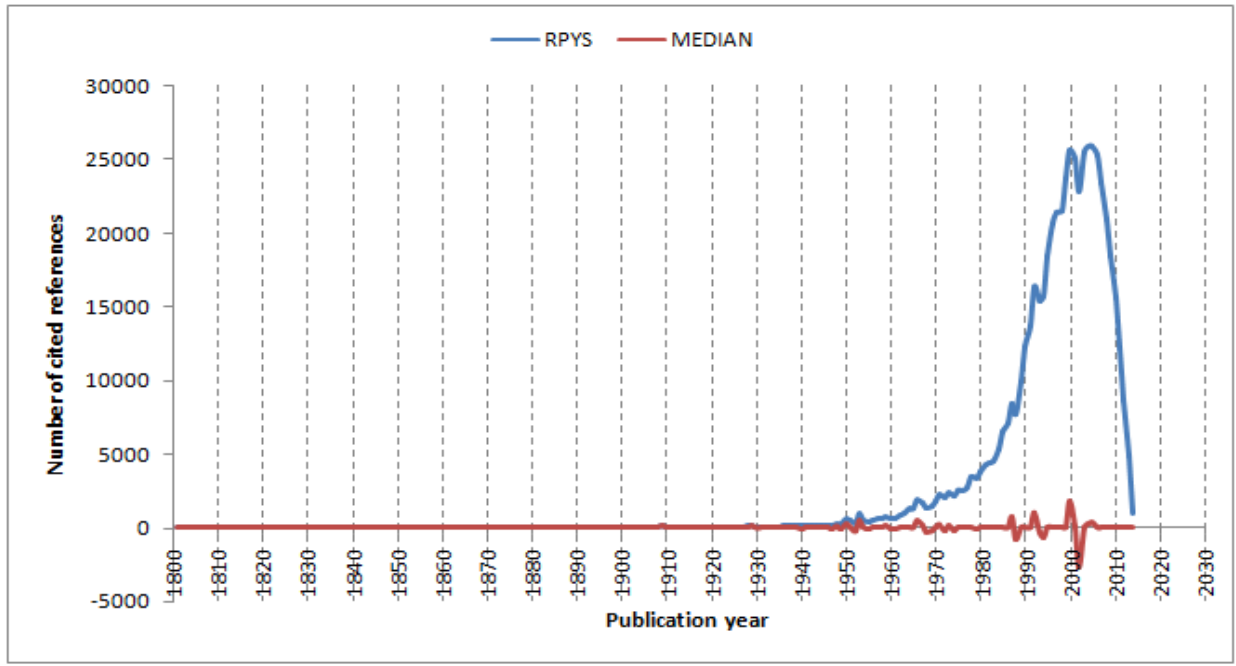

Figure 1 – Reference publication years (1801 – 2014) of tribology research publications (published between 1953 and 2014)

| Year | Q | Year | Q | Year | Q | Year | Q | Year | Q |
|---|---|---|---|---|---|---|---|---|---|
| 1801 | 8.41 | 1844 | 21.96 | 1887 | 17.29 | 1930 | 57.01 | 1973 | 81.31 |
| 1802 | 7.95 | 1845 | 11.22 | 1888 | 41.12 | 1931 | 56.54 | 1974 | 80.38 |
| 1803 | 16.82 | 1846 | 21.5 | 1889 | 8.88 | 1932 | 59.82 | 1975 | 82.25 |
| 1804 | 24.3 | 1847 | 10.75 | 1890 | 29.91 | 1933 | 59.35 | 1976 | 81.78 |
| 1805 | 50.47 | 1848 | 0.94 | 1891 | 45.8 | 1934 | 60.28 | 1977 | 82.71 |
| 1806 | 16.36 | 1849 | 21.03 | 1892 | 39.25 | 1935 | 60.75 | 1978 | 83.65 |
| 1807 | 7.48 | 1850 | 28.04 | 1893 | 51.87 | 1936 | 67.76 | 1979 | 83.18 |
| 1808 | 15.89 | 1851 | 0.47 | 1894 | 45.33 | 1937 | 66.36 | 1980 | 84.11 |
| 1809 | 15.42 | 1852 | 20.56 | 1895 | 42.99 | 1938 | 64.96 | 1981 | 84.58 |
| 1810 | 7.01 | 1853 | 36.45 | 1896 | 52.34 | 1939 | 66.82 | 1982 | 85.05 |
| 1811 | 6.54 | 1854 | 27.57 | 1897 | 35.05 | 1940 | 63.09 | 1983 | 85.52 |
| 1812 | 6.08 | 1855 | 32.25 | 1898 | 42.53 | 1941 | 62.62 | 1984 | 86.45 |
| 1813 | 5.61 | 1856 | 0 | 1899 | 34.58 | 1942 | 65.89 | 1985 | 86.92 |
| 1814 | 5.14 | 1857 | 20.1 | 1900 | 44.86 | 1943 | 62.15 | 1986 | 87.39 |
| 1815 | 14.96 | 1858 | 31.78 | 1901 | 47.2 | 1944 | 63.55 | 1987 | 88.32 |
| 1816 | 14.49 | 1859 | 10.28 | 1902 | 57.48 | 1945 | 64.49 | 1988 | 87.85 |
| 1817 | 4.68 | 1860 | 27.11 | 1903 | 40.66 | 1946 | 68.23 | 1989 | 89.25 |
| 1818 | 4.21 | 1861 | 31.31 | 1904 | 46.73 | 1947 | 65.42 | 1990 | 90.19 |
| 1819 | 3.74 | 1862 | 26.64 | 1905 | 44.39 | 1948 | 69.16 | 1991 | 90.66 |
| 1820 | 14.02 | 1863 | 41.59 | 1906 | 43.46 | 1949 | 68.69 | 1992 | 92.53 |
| 1821 | 34.11 | 1864 | 9.82 | 1907 | 51.4 | 1950 | 71.96 | 1993 | 91.12 |
| 1822 | 29.44 | 1865 | 26.17 | 1908 | 50.94 | 1951 | 71.5 | 1994 | 92.06 |
| 1823 | 3.27 | 1866 | 30.84 | 1909 | 64.02 | 1952 | 69.63 | 1995 | 92.99 |
| 1824 | 2.81 | 1867 | 35.98 | 1910 | 47.67 | 1953 | 76.17 | 1996 | 93.93 |
| 1825 | 33.65 | 1868 | 19.63 | 1911 | 42.06 | 1954 | 70.56 | 1997 | 94.86 |
| 1826 | 13.55 | 1869 | 19.16 | 1912 | 38.79 | 1955 | 70.1 | 1998 | 95.33 |
| 1827 | 2.34 | 1870 | 30.38 | 1913 | 49.07 | 1956 | 71.03 | 1999 | 96.73 |
| 1828 | 1.87 | 1871 | 18.69 | 1914 | 49.54 | 1957 | 72.43 | 2000 | 98.6 |
| 1829 | 23.83 | 1872 | 9.35 | 1915 | 38.32 | 1958 | 73.37 | 2001 | 97.2 |
| 1830 | 13.09 | 1873 | 40.19 | 1916 | 46.26 | 1959 | 74.3 | 2002 | 95.8 |
| 1831 | 28.97 | 1874 | 18.23 | 1917 | 36.92 | 1960 | 72.9 | 2003 | 98.13 |
| 1832 | 23.37 | 1875 | 35.52 | 1918 | 48.6 | 1961 | 73.83 | 2004 | 99.07 |
| 1833 | 33.18 | 1876 | 17.76 | 1919 | 48.13 | 1962 | 74.77 | 2005 | 99.54 |
| 1834 | 32.71 | 1877 | 25.7 | 1920 | 54.21 | 1963 | 75.24 | 2006 | 97.67 |
| 1835 | 28.51 | 1878 | 25.24 | 1921 | 55.61 | 1964 | 76.64 | 2007 | 96.26 |
| 1836 | 43.93 | 1879 | 37.39 | 1922 | 53.74 | 1965 | 77.11 | 2008 | 94.39 |
| 1837 | 12.62 | 1880 | 24.77 | 1923 | 56.08 | 1966 | 79.44 | 2009 | 93.46 |
| 1838 | 12.15 | 1881 | 58.88 | 1924 | 53.27 | 1967 | 78.51 | 2010 | 91.59 |
| 1839 | 1.4 | 1882 | 61.22 | 1925 | 57.95 | 1968 | 77.57 | 2011 | 89.72 |
| 1840 | 22.9 | 1883 | 52.81 | 1926 | 54.68 | 1969 | 78.04 | 2012 | 88.79 |
| 1841 | 11.68 | 1884 | 39.72 | 1927 | 58.41 | 1970 | 78.97 | 2013 | 85.98 |
| 1842 | 37.85 | 1885 | 50 | 1928 | 61.68 | 1971 | 80.84 | 2014 | 75.7 |
| 1843 | 22.43 | 1886 | 55.14 | 1929 | 67.29 | 1972 | 79.91 | | |

Figure 2 – Quantiles of the yearly number of cited references. The higher the quantile for a specific reference publication year, the darker the corresponding cell.

The RPYs are grouped into two periods of investigation, i.e. 1801-1900 and 1901-1965, based on the following remarks: (i) the 19$^{th}$ century was an era of rapidly accelerating scientific discovery and invention with significant developments in the fields of mathematics, physics, chemistry, biology, electricity, and metallurgy. The developments laid the groundwork for the technological advances of the 20$^{th}$ century. With the RPYs, we expect to find basic literature in the historical science which is also important for tribology. (ii) Since the term "tribology" was introduced in 1966 by Jost (1966), we study the period between 1901 and 1965 to identify early important publications for the field at the beginning and at the middle of the 20$^{th}$ century.

Reference Publication Years from 1801 to 1900

There are five larger peaks exhibited between 1801 and 1900 (in a span of 100 years). As the deviations from the median (red line) in Figure 3 shows, these peaks appear in 1805, 1882, 1886, 1893, and 1896. Obviously, some important historical papers for the development of tribology research were published at end of the 19$^{th}$ century.

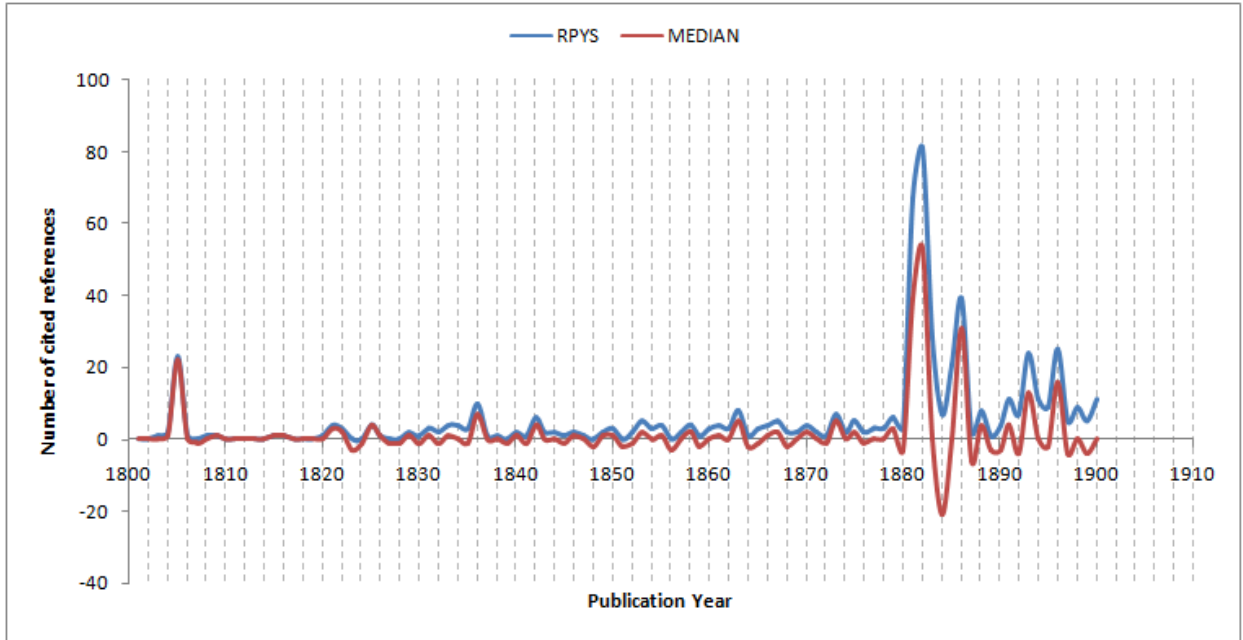

Figure 3 – Reference publication years between 1801 and 1900

If one analyzes the publications underlying RPYS peaks in the 19th and the first half of the 20th century, they often go back to single highly-cited publications (Marx et al. 2014). This is also the case in the current study, as shown by the results in Table 2.

| Table 2 - Most frequently cited publications between 1801 and 1900 | | |
|---|---|---|
| **RPY** | **TCR** | **Frequently Cited Publications** |
| 1805 | 23 | All refer to Young T (1805). PhilosTrSocLond, v95: p65 |
| 1882 | 82 | 78 refer to Hertz H (1882). Angenw Math, 92: 156. |
| 1886 | 39 | 34 refer to Reynolds O (1886). PhilosTrSocLond, 177: 157 |
| 1893 | 24 | 14 refer to Barus C (1893). Am J Sci, 45:87 |
| 1896 | 25 | 21 refer to Hertz H (1896). Miscellaneous Papers, 146 |
| RPY = Reference Publication Year, TCR = Total Number of Cited References | | |

The first peak in 1805 refers to the paper "An Essay on the Cohesion of Fluids" by **Young (1805)**. Contact angle and wetting are the starting points for heterogeneous thin film development. The concept of "surface tension" was also introduced in Young (1805). Young's equation describes the force balance between the interfacial tensions formed at the solid–liquid–vapor contact line. This equation is being used to calculate the surface tension and contact angle even now after centuries (Quere and Reyssat 2008; Simpson et al. 2015).

The second peak in 1882 refers to the paper "Über die Berührung fester elastischer Körper" (On the Contact of Elastic Solids) by **Hertz (1882)** published initially in German. Contact mechanics originated from Hertz's work, played an important role in tribology and other engineering applications. It provides necessary information for the safe and energy efficient design of technical systems and for the study of tribology and hardness of indentation. Hertz (1882) formulated the law of interaction which is a landmark in the field of linear elasticity. The Hertzian contact theory is being used to determine the relationship between contact pressure distribution and contact radius (Song and Gu 2012). Hertzian contact stress forms the foundation for the equations for load bearing capabilities and fatigue life in bearings, gears, and other bodies where surfaces are in contact.

The third peak in 1886 refers to the paper "On the Theory of Lubrication and Its Application to Mr. Beauchamp Tower's Experiments, Including an Experimental Determination of the Viscosity of Olive Oil" by **Reynolds (1886).** The author reveals classic examples on film lubrication. Reynolds' equation on film lubrication and pressure describes fluid flow accurately. This leads to various applications in dampers of aircraft, gas turbines, gear boxes, journal bearings, air bearings, and human joints in the usage of smooth surface geometrics of elastohydrodynamic lubrication.

The fourth peak in 1893 is especially based on the article "Isothermals, Isopiestics and Isometrics relative to Viscosity" by **Barus (1893)**. In this article, Barus provides a relationship between the viscosity and pressure of liquids. This is known as the Barus equation. Conventional viscometry normally uses the Barus equation for correlations. The viscosity-pressure dependence described by the well-known Barus law is extensively used by the engineers. Later, van Leeuwen (2009) proved the Barus equation to be non-applicable at high film pressures of 1 GPa or more.

The fifth peak in 1896 traces back to the article "On the contact of elastic solids" by **Hertz (1896).** It is the English translation of **Hertz (1882)**.

Reference Publication Years from 1901 to 1965

There are six larger peaks exhibited between 1901 and 1965 (in a span of 65 years). As the deviations from the median in Figure 4 show, these peaks appear in 1909, 1929, 1948, 1950, 1953, and 1959. The peaks suggest that important papers for the development of tribology research have been published in the 20$^{th}$ century before the term "tribology" was introduced in 1966. The papers which have been most frequently cited in the six peak years (see Figure 4) are listed in Table 3.

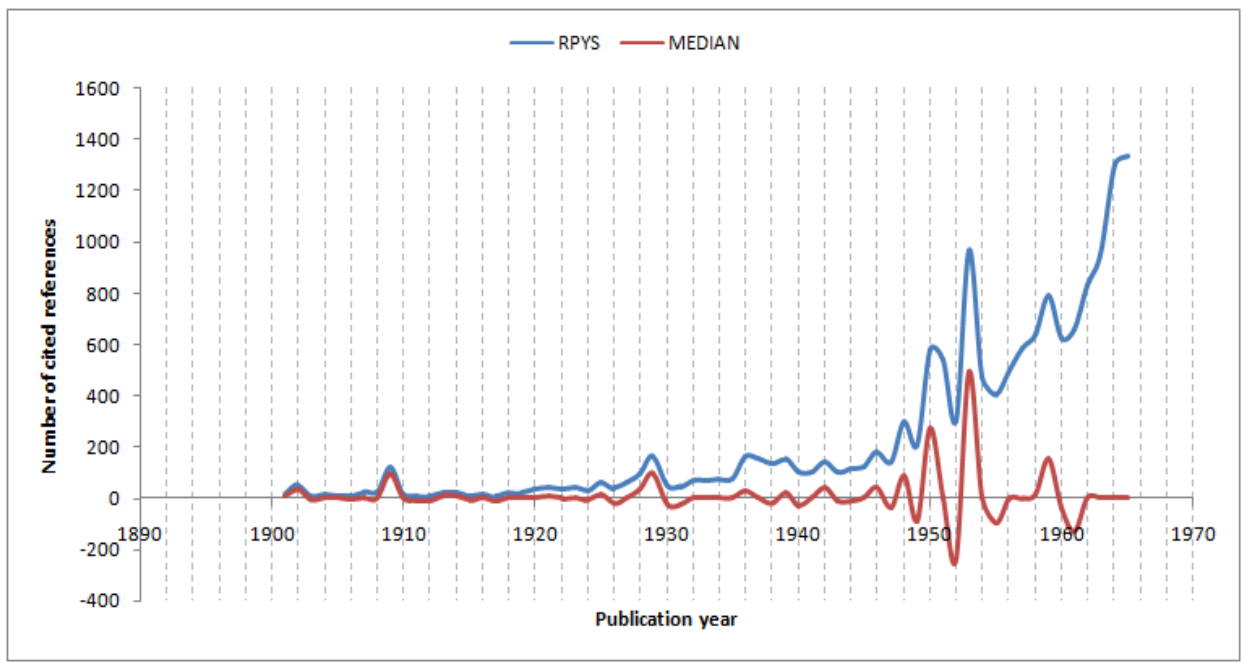

Figure 4 – Reference publication years between 1901 and 1965

| Table 3 – Most frequently cited publications between 1901 and 1965 |||
|---|---|---|
| **RPY** | **TCR** | **Frequently Cited Publications** |
| 1909 | 121 | 102 refer to Stoney G G (1909). P Roy ScoLond A, 82: 172. |
| 1929 | 163 | 107 refer to Tomlinson G A (1929). Philos Mag, 7: 905. |
| 1948 | 299 | 72 refer to Savage R H (1948). J App Phys, 19: 1. |
| 1950 | 579 | 233 refer to Bowden F P (1950). Friction Lubrication, 1/2. |
| 1953 | 968 | 484 refer to Archard J F (1953). J app Phys, 24: 981. |
| 1959 | 792 | 128 refer to Archard J F (1959). Wear, 2: 438. |
| RPY = Reference Publication Year, TCR = Total Number of Cited References |||

The first peak in 1909 refers to the article "The Tension of Metallic Films deposited by Electrolysis" by **Stoney (1909)**. Stresses in thin films are determined mainly using Stoney's equation which explains the relationship between the surface stress change and cantilever's tip deflection.

The second peak in 1929 is especially based on the article "A Molecular Theory of Friction" by **Tomlinson (1929)**. A pioneering attempt to explain friction on the atomic level was made in this article. Accordingly, friction is due to the interaction of molecules very close to each other which leads to the prediction of lattice properties and friction between various materials.

The third peak in 1948 goes back to the article "Graphite Lubrication" by **Savage (1948)**. Due to strong cohesion of planes, graphite becomes fine dust which leads to its failure of lubrication in vacuum as founded by Savage (1948).

The fourth peak in 1950 refers to the book "The Friction and Lubrication of Solids" by **Bowden and Tabor (1950)** which is an important landmark in the development of tribology research. David Tabor is the first recipient of the Tribology Gold Medal. The book covers the behavior of non-metals, especially elastomers, elastohydrodynamic lubrication, and the wear of sliding surfaces, which gradually replaced the earlier concept of the friction mechanism. The adhesion theory advocated by Bowden and Tabor is accepted as the fundamental theory of friction in the field of tribology.

The fifth peak in 1953 traces back to the article "Contact and Rubbing of Flat Surfaces" by **Archard (1953)**. Number and size of contact areas increase with the load on the model upon which mechanical wear and electrical contact also depend. Hence, high hardness of tool material maximizes the tool life as stated in Archard Wear Law used in sliding wear.

The sixth peak in 1959 is especially based on the article "The Temperature of Rubbing Surfaces" by **Archard (1959)** where a condensed version of flash theory is proposed. Later the theory became an idealized model in the rubbing contact.

**DISCUSSION**

RPYS implies to analyze the early RPYs cited within the body of publications of a specific research field. Major contributions (single frequently referenced publications) appear as prominent peaks in the time series regarding the frequency of cited references as a function of RPYs. As a rule, these contributions are the origins or historical roots of a research field (Barth et al. 2014). Recently the RPYS was used by Barth et al. (2014) in physics, by Leydesdorff et al. (2014) in information science, by Marx and Bornmann (2014) in biology, and by Comins and Hussey (2015) in global positioning systems.

In this study, the RPYS software (Marx et al. 2014; Bornmann et al. in press) is used to analyze the important historical publications in tribology research. The results on tribology show that RPYS has the potential to identify the important publications in the early history of tribology research: most of the publications which have been identified in this study as highly referenced (cited) publications are landmark publications in the field of tribology.

Even though, the term tribology was coined during 1966 by Jost (1966), the basic of tribology dates centuries back. Tribology started with the thin film development and contact mechanism initially. A subsequent development was Reynolds' equation which had led to various applications using fluid flow. The developed Barus equation is used only for fluid flow and viscosity at low pressure. The further development in tribology was initiated with friction theory and lubrication. During the mid of the $20^{th}$ century, tool lives were improved using wear law and contact friction. Further, wear mechanism maps played an important role.